\documentclass[a4paper,fleqn,useAMS,usenatbib]{mnras}

\pdfoutput=1

\usepackage{natbib}
\usepackage{mathptmx}

\usepackage[T1]{fontenc}
\usepackage{ae,aecompl}

\usepackage[dvips]{graphicx}
\usepackage{amsmath}
\usepackage{amsfonts}
\usepackage{amssymb}
\usepackage{comment}
\usepackage{bm} 
\usepackage[dvipsnames]{xcolor}
\usepackage[]{hyperref}
\hypersetup{
	colorlinks=true,	
	urlcolor=MidnightBlue,		
	hypertexnames=true,
	plainpages=false,
	naturalnames=true,
	pdftitle={Loss Wedge},    
	pdfauthor={Zephyr Penoyre},     
	linkcolor=WildStrawberry,          
	citecolor=Periwinkle,        
}

\usepackage[percent]{overpic}
\usepackage{subfig}
\usepackage{csvsimple}
\usepackage{lscape}
\usepackage[shortlabels]{enumitem}

\DeclareMathAlphabet{\mathcal}{OMS}{ntxsy}{m}{n}   
\SetMathAlphabet{\mathcal}{bold}{OMS}{ntxsy}{b}{n} 

\makeatletter
\def\env@matrix{\hskip -\arraycolsep 
  \let\@ifnextchar\new@ifnextchar
  \array{*{\c@MaxMatrixCols}c}}
\makeatother

\newcommand{\cz}{\mathrm{c}}
\newcommand{\sz}{\mathrm{s}}

\title[Blended point sources]{The position and resolvability of blended point sources}
\author[z. penoyre]{zephyr penoyre$^{1}$\thanks{E-mail:
\href{mailto:zephyrpenoyre@gmail.com}{zephyrpenoyre@gmail.com}}\\
$^{1}$Leiden Observatory, Leiden University, P.O. Box 9513, 2300 RA Leiden, the Netherlands}
\date{Submitted to RASTI 24 Apr 2025. Accepted 4 Dec 2025;}

\pubyear{2025}

\begin{document}
\label{firstpage}
\pagerange{\pageref{firstpage}--\pageref{lastpage}}
\maketitle

\begin{abstract}
In this work we derive analytic expressions and numerical recipes for finding the effective observed position of sources close enough on sky that their Point Spread Functions (PSF), modelled as Gaussian profiles, overlap. In particularly we derive these for an elongated PSF, with a long and short axis, such as we would see from an instrument with a rectangular or elliptical mirror (relevant, for example, for the \textit{Gaia} mission). We show that in this case the problem can be reduced to a one dimensional brightness profile with extrema along the line connecting the two sources, with an effective PSF width that depends on the relative orientation of the PSF and its degree of elongation. The problem can then be expressed in units of this effective width to be a function of the relative separation and light ratio alone (thus reducing to a rescaling of the un-elongated case). We derive the minimum light ratio, for a given separation and effective width, above which two sources will be resolved. We map out numerical procedures for finding the positions of these extrema across all possible cases. Finally we derive the positional offset and deviance associated with observing a fixed pair of blended sources from a variety of orientations, showing that this can be a significant source of excess noise.
\end{abstract}


\section{Introduction}

Fundamental physical limits on the diffraction of light cause any point-like source of light observed via a lens (or curved mirror) to produce an image of finite width. This finite Point Spread Function (PSF) has a characteristic angular width $\theta \sim \frac{\lambda}{d}$ where $\lambda$ is the wavelength of the light and $d$ the diameter of the lens/mirror.

Thus when we observe a distant physical object of size $R$ at distance $D$ ($\gg R$) we can resolve it only when $\frac{R}{D} \gtrsim \theta$ - otherwise it is unresolved and we see all the incident light spread over the PSF of width $\theta$.

The vast majority of visible stars (whether observed by eye or modern telescopes) fall well below this resolvable limit. A second star may reside close enough on-sky to the first (at a separation $r \lesssim D \theta$) that the two cannot be separately resolved and an image of the pair appears to be a single PSF with the blended properties of the two sources, visually indistinguishable from a single source.

This applies to pairs of sources that are truly close in physical space (e.g. perhaps a gravitationally bound binary system) or that are at very different distances from the observer and each other but appear close in projection. It is also eminently possible for more than two sources to be blended - for example whole clusters of stars appear as a bright point source in extragalactic observations \citep{Harris91}. In fact we may expect many blended sources to be contributing to an observed image (for example planets and asteroids around a star, or many more distant background stars) though if these systems contribute a negligible fraction of the observed light, their influence is small.

Any inference we might make about an observed source is altered by the knowledge that they are in fact multiple sources blended. For example we might hope to translate a measurement of a star's brightness and distance into an estimate of mass, but this often assumes a single star, or relies on knowing the fractional contribution of each source to the blended image.

Binary stars (and higher multiples) are ubiquitous in most astrophysical environments. Only very wide systems, or those very close to the observer, are resolvable - however multiplicity affects many facets of their behaviour and evolution. To some they are objects of interest in their own right, to others a nuisance to be removed or compensated for when making observations. Only a subset can be identified, and only a subset of those can have their properties measured. The number of known multiples is rapidly increasing, both by resolving wide systems (for example \citealt{Tokovinin24}) or, for unresolved systems, by identifying variable or excess motion (spectroscopically or astrometrically, see \citealt{El-Badry24} for a recent review) or light (e.g. \citealt{Prsa05} or \citealt{Wallace24}).

The \textit{Gaia} survey has been crucial to the improvement in our capacity to detect stellar multiples, as well as greatly expanding our knowledge of the position and motion of stars in the Milky Way \citep{Prusti16,Vallenari23}. It is, at least in part, an astrometric survey - scanning repeatedly over the sky and making many measurements of the positions of sources, from which their motion can be inferred \citep{Lindegren21}. In \citet{Penoyre20} we mapped out how the orbits of unresolved binaries contribute to this motion, due to the offset between their photocentre (or center-of-light) and their center-of-mass. In appendix A of that work, we derived the simple first order photocentre offset distance, without including an important nuance to \textit{Gaia}'s functionality: that its mirror is rectangular, measuring 1.45 m by 0.5 m \citep{Prusti16}, and hence its subsequent PSF is extended along the direction corresponding to the short axis of the mirror.

In this paper we update this calculation, identifying when this nuance significantly alters observations of a system. In section \ref{sec:coord} we find the brightness profile of an unresolved system given an elongated PSF. In section \ref{sec:resolvability} we go on to find analytically the conditions for such a brightness profile to be resolvable as two sources, and the position of the peak (or peaks). This applies equally to any close pair of sources on sky (regardless of whether they are gravitationally bound) and thus is as appropriate for modelling blending with background sources as for unresolved multiple systems. Finally in section \ref{sec:obs} we focus on the case of a pair of sources in a fixed relative configuration measured over a range of observing angles, appropriate for long period binaries, distant sources, or chance pairs that move in the same fashion. We find the positional offset and excess noise that would be inferred when fitting these as a single source.

\section{Brightness profiles}
\label{sec:coord}

Throughout this work we will work with Gaussian point-spread functions (PSF). These are vastly simplified from real diffraction patterns. Idealised forms of elongated PSFs include convolutions of two $\mathrm{sinc}^2$ functions for a rectangular aperture, or an elongated form of the well known Airy disk for an elliptical aperture \citep{Airy35,Raman19}. 

Idealised PSFs are defined in terms of a single wavelength, but detectors are sensitive to a range of wavelengths (which may be quite broad, 330-1000 nm in \textit{Gaia} for example). The image of such a source will be the convolution of each PSF with the sources apparent brightness at that wavelength.

Idealised diffraction patterns generally have secondary (and higher) maxima of lower amplitude. Though these may be smoothed out by convolving over wavelength, or physical limitations of the instrument, this still leads to a small but not insignificant amount of brightness at large separations from the center of the image.

True PSFs will be further altered by any non-ideal aberrations to the mirror, optical path, or detector. For example \citet{Fabricius16} show a typical Gaia PSF (figure 7) which as expected resembles a smoothed $\mathrm{sinc}^2$ function, but displays significant asymmetry and strong diagonal spikes. Real observations are then further altered by pixelation, losing more information as the incident flux is binned over a finite number of pixels. Very bright sources can oversaturate these pixels, spilling charge and thus apparent brightness into neighbouring pixels, and further deforming the PSF.

In contrast a Gaussian profile has a single central maxima, and diminishing brightness at large separations. It is thus generally an imperfect proxy for a true PSF. It is however a reasonable approximation for the central core of the PSF, which as we will show dominates our resolvability criteria. It also has many mathematical conveniences, including the fact that any slice of an elongated 2D Gaussian profile is itself a Gaussian profile.

\subsection{Imaging close sources}

\begin{figure*}
\includegraphics[width=0.98\textwidth]{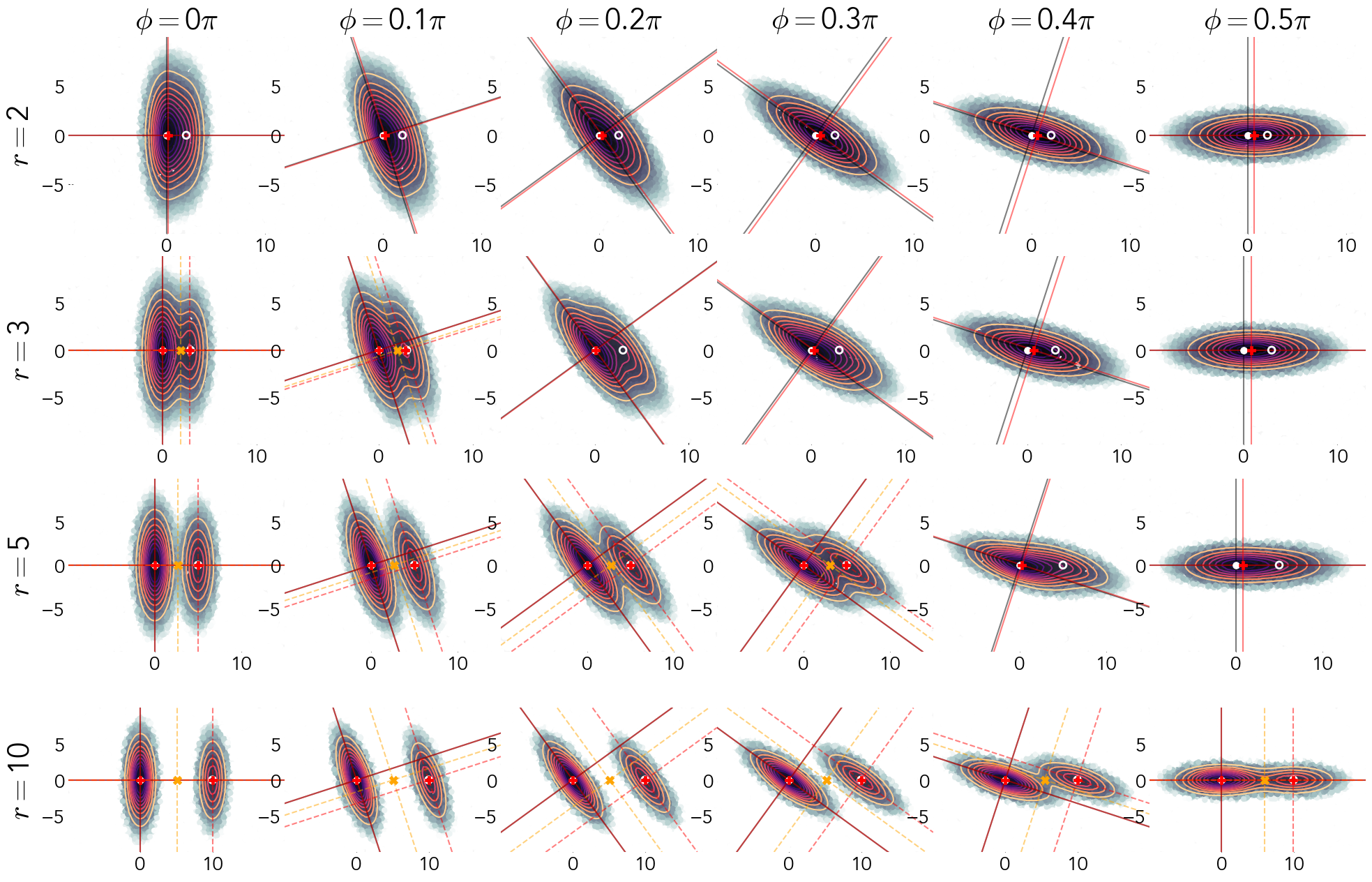}
    \caption{A grid of example images (as given by $b(x,y)$, equation \ref{eq:b_xy}) of two sources with light-ratio $l=0.5$ observed at different separations $r$ (rows) and angles $\phi$ (columns) by an instrument with an elongated PSF with relative width $\beta=3\alpha$. We work in coordinates $x,y$, on-sky distances are in units of $\alpha$, with the primary at the origin and the secondary on the line $y=0$. The two white circles show the positions of the two point sources. Red $+$ symbols denote maxima of the brightness profile, and where present orange $\times$ symbols denote a saddle-point (in which cases there are two maxima). The red-orange-yellow contour lines show 10 contours of equal brightness evenly spaced between the maximum value and 0. Straight lines show the along- and across- scan positions of the primary (black), primary maximum (red), and where present the saddle-point (dashed orange) and secondary maximum (dashed red). Unresolved systems (no orange $\times$ or line) occur at low separation, but also at larger separations if the PSF is oriented such that the long axis is close to parallel with the line separating the pair.}
    \label{fig:col_grid}
\end{figure*}

Let us start by imagining two sources separated by an on-sky distance\footnote{As this paper focusses only on observed systems on-sky we'll use $x$, $y$ and $r$ as on-sky distances (measured as an angle). For a true instantaneous physical distance $R$ and a separation aligned at angle $\theta$ from the line of sight $r=\varpi \frac{R}{\mathrm{AU}} \sin(\theta)$, and the unit of $r$ is the same as that which is used for parallax $\varpi$.} $r$. As we're imagining systems where $r$ is relatively small, they may be close in physical space (e.g. a binary) or far apart and simply close in projection on-sky (e.g. blended background and foreground stars).

We imagine that both sources are observed in a single image with an instrument that has an elongated Gaussian PSF of width $\alpha$ in the short direction and $\beta \ (>\alpha)$ in the long direction.

For derivations in this paper we'll work entirely in units of $\alpha$ and $\beta$ (and hence we shall not need numerical values specific to a given instrument). However, as \textit{Gaia} is the most currently relevant point of comparison we note that its full-width-half-maximum is approximately 0.1 arcseconds \citep{Fabricius16} in the along scan direction, and hence $\alpha \sim 0.04$ arcseconds. Given that the ratio of side lengths of the mirror is 1.45:0.5 we expect $\beta \sim 3 \alpha$.

If $r \gg \alpha,\beta$ then we expect the two sources to be clearly \textit{resolved} - but the case of interest to us is where $r$ is sufficiently small that the two PSFs overlap significantly. For a system of fixed physical size and an instrument of fixed PSF width this is equivalent to observing from a large enough distance (small enough parallax, $\varpi$).

In the \textit{unresolved} limit ($r \ll \alpha,\beta$) the image will appear to be approximately that of a single source. As $r$ increases it will start to noticeably deform, and around $r \sim \alpha,\beta$ (a transition we will quantify more carefully later in this text) the two peaks will be on the edge of being resolvable seperately - a regime we will call \textit{semi-resolved}.

We are interested in the small offset of the inferred position of the pair in the unresolved limit, how this behaves as we approach the semi-resolved regime, and at what point resolving the two sources separately becomes possible.

To approach this algebraically we will make two strong assumptions about what measurements we would infer for such a system, based on the \textit{brightness profile} of the image convolved with the PSF. These assumptions are that
\begin{itemize}
\item an unresolved pair will be measured to have a position corresponding to the maximum brightness 
\item a pair of sources is resolved when there is a minimum in the 1D brightness profile (or saddle point in 2D) between the two sources, and correspondingly two distinct maxima
\end{itemize}

Real observations may use a more complex PSF fitting technique (e.g. \citealt{Rowell21,Harrison23}), but the above assumptions provide a reasonable approximation that is well defined mathematically.

We can label the two sources $A$ and $B$ and can define a light ratio between the two
\begin{equation}
l=\frac{F_B}{F_A}
\end{equation}
where $F_{I}$ is the flux observed from source $I$. We take $l\le 1$ always, such that the primary, $A$, is the object that we are more able to see.

We can work in coordinates $x,y$ (which we will take to be on-sky angles), aligned with the separation of the two sources and centred on $A$, such that $\mathbf{r}_A = (x_A,y_A) = (0,0)$ and $\mathbf{r}_B = (r,0)$.

Let us take the alignment of our instrument (and hence the orientation of the PSF) to be at a \textit{scan angle}, $\phi$, relative to the vector from the $A$ to $B$.

The observed brightness profile of one source, at a position $\Delta \mathbf{r} = (\Delta x,\Delta y)$ from its center is thus
\begin{equation}
\label{eq:Esingle}
E(\Delta x,\Delta y) = e^{-\frac{\left(\cz_\phi \Delta x + \sz_\phi \Delta y \right)^2}{2\alpha^2}}e^{-\frac{\left(-\sz_\phi \Delta x + \cz_\phi \Delta y \right)^2}{2\beta^2}}
\end{equation}
(using the shorthand $\sz_x =\sin(x)$ and similarly for $\cos$). The first term is the result of projection of the offset position along the short axis, and the second is due to the projection along the long axis.

Then the combined brightness of the pair (up to an arbitrary rescaling) is
\begin{equation}
\label{eq:b_xy}
b(x,y) = E(x,y) + l E(x-r,y).
\end{equation}

\subsection{Maxima and minimum}

Any extrema in $b(x,y)$ occurs at a position $\mathbf{r}_m =(x_m,y_m)$ such that $\partial_x b|_{\mathbf{r}_m}\  =\ \partial_y b|_{\mathbf{r}_m} = 0$.

Returning to the single source profile (equation \ref{eq:Esingle}) it can clearly be seen that it peaks at $\Delta x= \Delta y =0$. Thus if the two sources are well separated ($r \gg \alpha,\beta$, such that at the location of one source, the contribution of the other is negligible) there will be two maxima. In contrast, if they are very close ($r \ll \alpha,\beta$, such that both contribute significantly to the brightness at or near their individual peaks) we will have one single maximum somewhere between the two. 

A series of example images is shown in figure \ref{fig:col_grid} which show both unresolved and resolved systems depending on separation and scan angle. Crucially, a system with constant $r$ can be resolved or unresolved depending on $\phi$ - specifically if the long axis of the PSF is close to aligned with the direction of separation.


To find the derivatives of $b$ we can first find
\begin{equation}
\partial_x E(\Delta x,\Delta y) = - \left(\frac{\Delta x}{\gamma^2} + \frac{\Delta y}{\nu^2}\right)E(\Delta x,\Delta y)
\end{equation}
and
\begin{equation}
\partial_y E(\Delta x,\Delta y) = - \left(\frac{\Delta x}{\nu^2} + \frac{\Delta y}{\omega^2}\right)E(\Delta x,\Delta y)
\end{equation}
where
\begin{equation}
\frac{1}{\gamma^2} = \frac{\cz_\phi^2}{\alpha^2} +\frac{\sz_\phi^2}{\beta^2}, 
\end{equation}
\begin{equation}
\frac{1}{\omega^2} = \frac{\sz_\phi^2}{\alpha^2} +\frac{\cz_\phi^2}{\beta^2},
\end{equation}
and
\begin{equation}
\frac{1}{\nu^2} = \sz_\phi \cz_\phi \left(\frac{1}{\alpha^2} -\frac{1}{\beta^2}\right), 
\end{equation}

A useful way to understand $\gamma$ and $\omega$ is to note that they are the width of the ellipsoid projected along the $x$-axis and $y$-axis respectively.

With these we can find
\begin{equation}
\partial_x b = -E(x,y)\left[ \left(1+\frac{l}{\Lambda}\right)\left(\frac{x}{\gamma^2} + \frac{y}{\nu^2}\right) - \frac{l}{\Lambda} \frac{r}{\gamma^2}\right]
\end{equation}
and
\begin{equation}
\partial_y b = -E(x,y)\left[ \left(1+\frac{l}{\Lambda}\right)\left(\frac{x}{\nu^2} + \frac{y}{\omega^2}\right) - \frac{l}{\Lambda} \frac{r}{\nu^2}\right]
\end{equation}
where
\begin{equation}
\Lambda(x,y) = \frac{E(x,y)}{E(x-r,y)}
\end{equation}
is the ratio of the PSF brightnesses, ignoring the relative flux, at some position $x,y$.

As $E(x,y)>0$ always, the condition $\partial_x b=0$ is simply that the square-bracket term be 0 (and equivalently for $\partial_y b=0$). These conditions are satisfied when
\begin{equation}
y_m=0
\end{equation}
and
\begin{equation}
\label{eq:x0}
x_m=\frac{l r}{\lambda_m + l}
\end{equation}
where $\lambda_m = \Lambda(x_m,0)$.

It is perhaps unsurprising that $y_m=0$, i.e. that the extrema necessarily lie on the line between the two sources - as both PSFs clearly decline for $|\Delta y|>0$. 
From now on we will only consider directly points on the line $y=0$ and thus define the slightly simplified
\begin{equation}
\label{eq:bx}
b_x(x)=b(x,0)= e^{-\frac{x^2}{2 \gamma^2}}\left( 1+ \frac{l}{\lambda} \right)
\end{equation}
and
\begin{equation}
\label{eq:etax}
\lambda(x)=\Lambda(x,0)= e^{\frac{r(r-2 x)}{2 \gamma^2}}.
\end{equation}

\begin{figure}
\includegraphics[width=0.98\columnwidth]{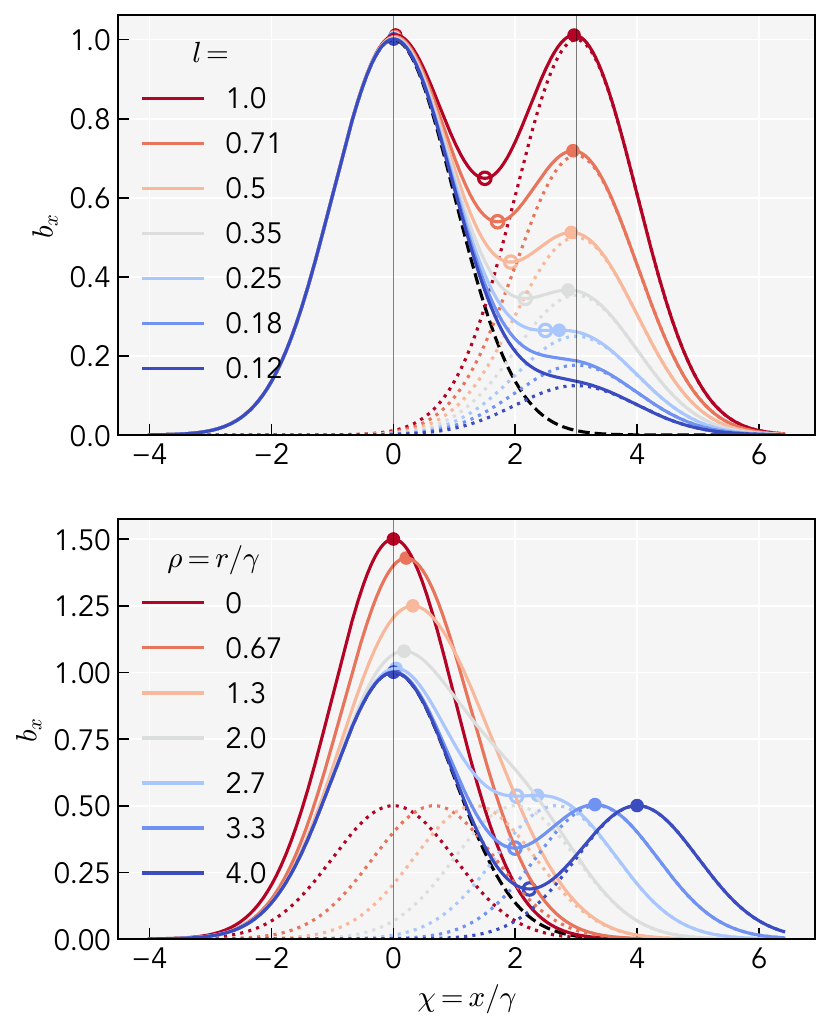}
    \caption{One-dimensional brightness profiles of two sources along their connecting line ($y=0$ and thus $b(x,y) \Rightarrow b(x,0) = b_x(x)$, equation \ref{eq:bx}). We work in units of the effective width, $\gamma(\phi)$ (see equation \ref{eq:gamma}), $\rho=r/\gamma$ and $\chi=x/\gamma$, and thus are able to remove any $\phi$ dependence from the functional form. The top panel shows sources of fixed $\rho=3$ and varying $l$ and the bottom panel shows sources of fixed $l=0.5$ and varying $\rho$. The black dashed line shows the profile of the primary (always fixed), coloured dotted lines show the profile of the secondary, and solid coloured lines the total brightness. Maxima are denoted by full circles and minimum (where present) by an open circle. Black vertical lines show the position of the primary, and in the first panel also the (fixed) position of the secondary.}
    \label{fig:col_bx}
\end{figure}

We can see that now the effective width term that sets the behaviour of the source is the orientation dependant $\gamma$. We can rewrite this in the more direct form
\begin{equation}
\label{eq:gamma}
\gamma(\phi) = \frac{\alpha}{\sqrt{1+\left(\frac{\alpha^2}{\beta^2}-1\right)\sz_\phi^2}}.
\end{equation}
This takes values $\alpha \leq \gamma \leq \beta$, with $\gamma=\alpha$ in the aligned ($\phi=0$ or $\pi$) case and $\gamma=\beta$ in the perpendicular ($\phi =\pm \frac{\pi}{2}$) case. 

We visualize $b_x$, the much simpler 1D brightness profile, in figure \ref{fig:col_bx}. Here we work in units of $\gamma$ which allows us to remove the dependence of angle $\phi$ from the profiles (though it is still important in setting the overall scale, and thus observed separations). Starting with the top panel, varying $l$ for fixed $\rho=r/\gamma$, we see that a minimum (corresponding to a saddle-point in the 2D image, and thus a resolved system) can exist for larger $l$, effectively disappearing at the moment the minimum and secondary maximum meet. In the second panel, varying $\rho$ at fixed $l$, we again see unresolved systems at smaller separations. Also can see more clearly now that the largest photocentre offset in unresolved pairs occurs at intermediate (small) $\rho$, here at $\rho \sim 1$, and reduces at larger $\rho$ even before the system becomes resolvable. In appendix \ref{ap:max_offset} we extend these calculations to find the conditions which give the maximum possible offsets.

Equations \ref{eq:bx} and \ref{eq:etax} are the equivalent of equation 64 in \citet{Lindegren22}. However, their calculation only includes the along-scan PSF width, $\alpha$ ($u$ in their notation), implicitly assuming $\beta \gg \alpha$ (and hence $\gamma \rightarrow \alpha/|\cz_\phi|$). This significantly overestimates the effective width as $|\cz_\phi|\rightarrow 0$.

\subsubsection{Approximate forms}

As $\lambda_m$ is a function of $x_m$, equation \ref{eq:x0} cannot be solved analytically. We can however see that as $\lambda_m \rightarrow \infty$, $x_m \rightarrow 0$ (i.e. a peak at the primary) and as $\lambda_m \rightarrow 0$, $x_m \rightarrow r$ (i.e. a peak at the secondary). These solutions only exist if $\lambda$ can take such extreme values in the range $0<x_m<r$. 

As $\frac{r}{\gamma} \rightarrow 0$, $\lambda \rightarrow 1$ and 
\begin{equation}
x_m \rightarrow x_0=\frac{lr}{1+l}
\end{equation}
where $x_0$ is the much simplified zeroth order equation for the photocentre reported in \citet{Penoyre20}. 

It is only in the marginally unresolved ($r \lesssim \gamma$) case where $\lambda_m$ approaches order unity and $x_m$ is significantly affected by $\gamma$. In appendix \ref{ap:approx} we derive the a correction accurate up to fourth order (equation \ref{eq:xm_simple}):
\begin{equation}
x_m = x_0 \left(1-\frac{x_0^2}{l\gamma^2}\right) + \mathcal{O}\left(\frac{r^4}{\gamma^4}\right).
\end{equation}

\section{Resolvability of sources}
\label{sec:resolvability}

\begin{figure}
\includegraphics[width=0.98\columnwidth]{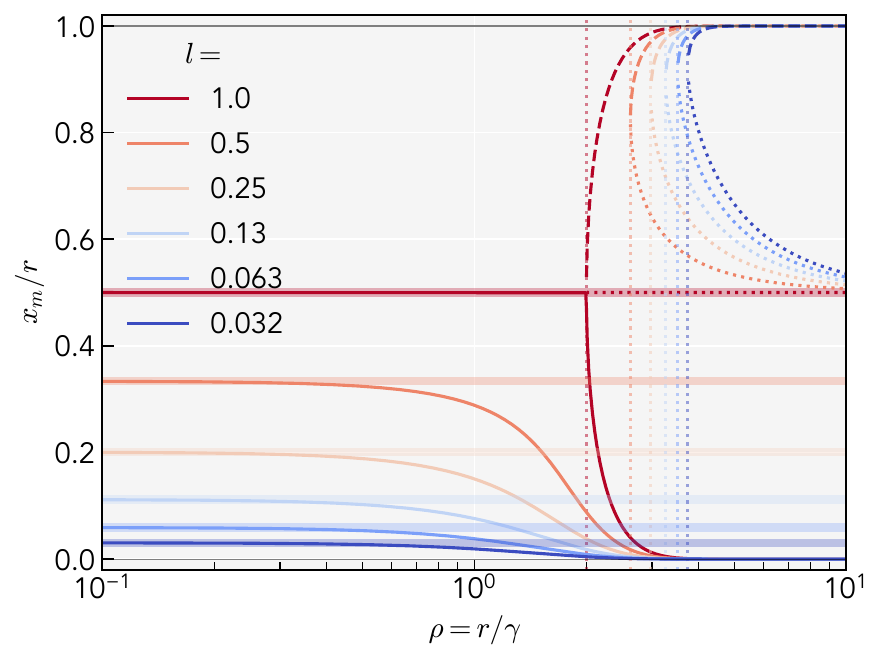}
    \caption{Here we show, as a function of $\rho$, the relative  location of the primary maxima (solid line) and where present the secondary maximum (dashed line) and the minimum (dotted line), for various $l$. The dashed vertical lines show the critical $\rho_c(l)$ above which a minimum exists. The wide translucent horizontal lines show $x_m = x_0$, the asymptotic solution for the position of the primary maximum when $\rho \ll 1$.}
    \label{fig:col_x}
\end{figure}

Now we will work in relative separations
\begin{equation}
\rho=\frac{r}{\gamma} \ \mathrm{and} \ \chi=\frac{x}{\gamma}
\end{equation}
which allows us to express all relevant expressions independently of $\gamma(\phi)$. The on-sky positions of any point of interest do scale with $\gamma$, but this need only be reintroduced at the end of the calculation. Similarly there is a nuance introduced in that pairs with a fixed $\rho$ do not have a fixed or independent separation or orientation. Nevertheless, it is very convenient that all the form of a brightness profile can be parametrised (and solved) in terms of $\rho$ and $l$ only.

We can return to equation \ref{eq:x0} but now express it in a way more amenable to numerical solution. We can define
\begin{equation}
\label{eq:g_chi}
g(\chi|\rho,l) = \chi - \frac{l\rho }{l+\lambda}
\end{equation}
where
\begin{equation}
\label{eq:lambda_chi}
\lambda(\chi|\rho) = e^{\frac{\rho(\rho-2\chi)}{2}}.
\end{equation}

Extrema in the brightness profile occur at zeros of $g$. These expressions can immediately (but not trivially) be solved numerically for $g=0$, but instead we'll explore the space of possible solutions analytically first to give more insight into the behaviour.

If $\chi$ is large and positive, $\lambda \rightarrow 0$ and $g \rightarrow \chi$ and thus $g$ is positive. Large negative $\chi$ gives $\lambda \rightarrow \infty$ and again  $g \rightarrow \chi$ and thus $g$ is negative. Thus we can see that there's always at least one real solution, but depending on  $l$ and $\rho$ there may be up to three real solutions.

A single real solution corresponds to a single maximum, whilst three real solutions corresponds to two maxima with a minimum between them (a saddle point in the 2D image). The limiting case, when there are two real solutions, corresponds to a repeated larger root and marks the transition from an unresolved pair (one maximum) to a resolvable pair (two maxima).

We can rearrange the previous expressions to write 
\begin{equation}
\label{eq:chi_lambda}
\chi = \frac{\rho}{2}-\frac{\ln \lambda}{\rho}.
\end{equation}
and
\begin{equation}
\label{eq:g_lambda}
g = \frac{\rho}{2}\left( \frac{\lambda-l}{\lambda+l} -\frac{2 \ln \lambda}{\rho^2}\right).
\end{equation}
This form is particularly useful if we consider the sign of $\ln \lambda$ (remembering $\lambda \ge 0$ for all $\chi$), which is positive for $\chi<\frac{\rho}{2}$, 0 at $\frac{\rho}{2}$, and negative for larger $\chi$.


For $\ln \lambda \geq 0$ ($\chi\leq \frac{\rho}{2}$) it is always possible to find a $\lambda \leq l$ such that $g=0$ - this corresponds to the ever present maximum near the primary. For $\ln \lambda < 0$ the right hand term in the bracket of equation \ref{eq:g_lambda} is positive, and for any root the left hand term must be negative - i.e. more solutions may exist for $0<\lambda<l$. For any given $\lambda$ the LHS term is larger (negative) for larger $l$, and the RHS term smaller (positive) for larger $\rho$ - thus as we have already seen (c.f. figure \ref{fig:col_bx}) multiple real solutions only occur at larger $l$ and $\rho$.

A necessary but not sufficient criterion for multiple maxima is that $g$ has a turning point, i.e. $\partial_\chi g \le 0$ for some $\chi$ (or equivalently $\partial_\lambda g \geq 0$ for some $\lambda$).
Using $\partial_\chi \lambda = -\rho \lambda$ we can find
\begin{equation}
\partial_\chi g = 1 - \frac{l \lambda}{(l+\lambda)^2}\rho^2
\end{equation}
which tends to 1 for large $|\chi|$.

Defining the critical point at which this zero of the derivative exists as $\chi_c$ such that $\partial_\chi g|_{\chi_c}=0$ we can find, in terms of the corresponding $\lambda_c=\lambda(\chi_c)$, the quadratic solution
\begin{equation}
\label{eq:lambda_0}
\lambda_c=l\left( \frac{\rho^2}{2}\left(1 - \sqrt{1-\frac{4}{\rho^2}}\right) - 1\right).
\end{equation}
The larger $\chi_c$ (smaller $\lambda_c$) solution is always the minimum of $g$ and thus we discard the positive root.
Real solutions only exist for $\rho \ge 2$ - i.e. double peaked brightness profiles can \textit{only} occur for $r \ge 2 \gamma$ (though again this is a necessary and not sufficient criterion; whether a minimum is visible for a given $\rho\geq 2$ depends on $l$).

For any $\rho \geq 2$ there is a critical minimum $l_c(\rho)$ (or equivalently a $\rho_c(l)$ for any $l$) below which double-peaked profiles cannot occur. This occurs when $g(\lambda_c)=0$ (i.e. the turning point in $g$ occurs at the root).

Two trivial cases exist: 
\begin{itemize}
\item when $\rho=2$, $\lambda_c=l$ which satisfies $g(\lambda_c)=0$ when $l=1$ (i.e. $l_c(2)=1$) showing that the closest pair that can be resolved are equally bright

\item as $\rho \rightarrow \infty$, $\lambda_c \rightarrow \frac{l}{\rho^2} \ll l$ and thus $g(\lambda_c) \rightarrow \frac{\rho}{2}\left(-1 - \frac{2 \ln \left(\frac{l}{\rho^2}\right)}{\rho^2}\right)$ which equals 0 for $l_c(\rho \rightarrow \infty) \rightarrow \rho^2 e^{-\frac{\rho^2}{2}}$ showing that at very large distances the secondary must be much dimmer to be unresolved (less bright than the PSF wings of the primary)
\end{itemize}

For any $l \geq l_c(\rho)$ (or $\rho \geq \rho_c(l)$) there exists a minimum of the brightness profile between $\chi=\frac{\rho}{2}$ and $\frac{\rho}{2} \leq \chi_c \leq 1$ (which can be found by substituting equation \ref{eq:lambda_0} into \ref{eq:chi_lambda}) and a secondary maximum between $\chi_c \leq \chi \leq \rho$.

We can summarise all this using the shorthand
\begin{equation}
\kappa(\rho) = \frac{\rho^2}{2}\left( 1- \sqrt{1-\frac{4}{\rho^2}}\right)
\end{equation}
such that $\lambda_c = l(\kappa -1)$. $\kappa$ takes values between 2 at $\rho=2$ and goes as $1+\frac{1}{\rho^2}$ as $\rho \rightarrow \infty$ (thus $1<\kappa\leq 2$). The brightness profile is double-peaked if and only if $\rho \geq 2$ (such that $\kappa$ has real values) and
\begin{equation}
g_c = g(\lambda=\lambda_c) = \frac{\rho}{2}\left( \frac{\kappa-2}{\kappa} - \frac{2 \ln \left(l(\kappa-1) \right)}{\rho^2}\right) \leq 0
\end{equation}
which can be rearranged to find explicitly the minimum light ratio which will give a double peaked profile for a given $\rho$,
\begin{equation}
\label{eq:l_c}
l_c(\rho) = \frac{1}{\kappa-1}e^{\frac{\rho^2}{2}\left(\frac{\kappa-2}{\kappa}\right)}.
\end{equation}

Using $\kappa$ we can also write down a shorthand form for $\chi_c$ of
\begin{equation}
\label{eq:chi_c}
\chi_c = \frac{\rho}{2}\left( 1 - \frac{\ln\left(l(\kappa-1)\right))}{\rho^2} \right)
\end{equation}
where we note that the value of the logarithm is always $\leq 0$ and thus $\chi_c \ge \frac{\rho}{2}$ as expected.

$l_c(\rho)$ cannot be easily inverted to find $\rho_c(l)$, which is a shame as this is perhaps the more useful form (imagine two sources of fixed brightness moving relative to each other) but the inversion can be done numerically.

\subsection{Numerical solutions}
\label{sec:numerical_solutions}

For a given system with known $l$ and $\rho$, we can immediately calculate $l_c(\rho)$ (equation \ref{eq:l_c}) and $\chi_c(\rho,l)$ (equation \ref{eq:chi_c}).

We seek the solution(s), $x_m$, of equation \ref{eq:x0}. This can be expressed as root(s) of $g(\chi_m)=0$ which can be reliably found by any root-solver which operates over a range $a < \chi < b$ such that $g(a)$ and $g(b)$ have opposite sign and there is a single root in this range (for example the van Wijngaarden-Deker-Brent method, \citealt{Brent02} - implemented as \texttt{brentq} in \texttt{scipy.optimize}).

In all cases there exists a first root near the primary, at on-sky position $x_1$, with $0<\chi<\frac{\rho}{2}$ - noting that $g(0) \leq 0$ (with equality when $l=0$) and $g(\frac{\rho}{2}) \geq 0$ (with equality when $l=1$) always.

If $l > l_c(\rho)$ there exists two more roots - an interstitial minimum, at $x_{min}$, and a maximum near the secondary source, at $x_2$. In this case $g(\chi_c)<0$ and the minimum lies in the range $\frac{\rho}{2} \leq \chi < \chi_c$ (with equality when $l=1$). $g(\rho) > 0$ always and thus when it exists the secondary maximum can be found in the range $\chi_c < \chi < \rho$.

In the case $l=l_c(\rho)$ both the minimum and secondary maximum coincide at a stationary inflection point at $\chi_c$ (and hence we don't need to resort to numerical root finding). 

The other problem case occurs when $l=1$ (and to some extent $l=0$, but this is a trivial case) - in this case $g(\frac{\rho}{2})=0$ and some root finders will fail. In this case the minimum (if it exists, $\rho \geq 2$) is trivially at $\chi=\frac{\rho}{2}$ - and the other two roots can be found by taking the limits of the range to be $\frac{\rho}{2} \pm \epsilon$ (where $\epsilon \ll \rho$ is some minuscule increment to avoid the 0 point). If there is no minimum ($l<l_c$) the maximum is situated at $\chi=\frac{\rho}{2}$. If $l=1=l_c$ (which implicitly means that $\rho=2$) all three turning points coincide at $\chi=\frac{\rho}{2}$, effectively giving a single maximum at that point.

The on-sky position, $x_m$, of any extrema can be found from the numerical solution $\chi_m$ by multiplying back through by the effective width $\gamma$.

\begin{figure}
\includegraphics[width=0.98\columnwidth]{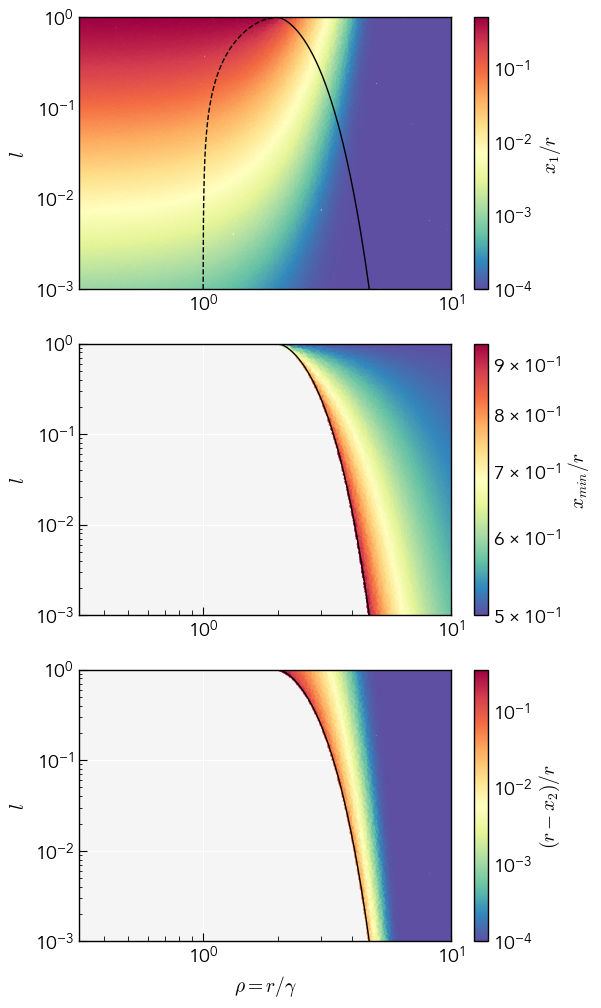}
    \caption{Numerically derived positions of the primary maximum, $x_1$, and where they exist ($l>l_c(\rho)$) the minimum $x_{min}$ and secondary maximum $x_2$. Here we show all quantities in units of $r$. The solid black line corresponds to the critical point beyond which three solutions (two peaks and a minima) exist, as described by equations \ref{eq:l_c} and \ref{eq:chi_c}). The dashed black line shows the $\rho$ that maximises $x_1$ for a given $l$, as detailed in appendix \ref{ap:max_offset}.}
    \label{fig:col_xs}
\end{figure}

The numerically derived position of each root is shown for a variety of $l$ and $\rho$ in figure \ref{fig:col_xs}. For small separations ($\rho<1$) the primary maximum is located close to the first order solution as found in \citet{Penoyre20}: $x_m \sim x_0$. As $\rho$ increases the relative offset decreases, fastest for smaller $l$ - thus the  first order solution is a good approximation for small $\rho$ and an overestimate for $\rho \gtrsim$ 1. In appendix \ref{ap:max_offset} we derive the conditions that maximise $x_1$, as is also shown in the figure. At low $l$ the photocenter offset is maximised at $\rho=1$, and as $l$ increases this goes to $\rho=2$.

We can clearly see the line ($l>l_c(\rho)$) above which minima and secondary maxima exist. Close to this boundary (particularly for low $l$) the minima occur very close to the secondary source and at larger $\rho$ it tends to the middle point ($x_{min} \rightarrow \frac{r}{2}$). Near the $l\geq l_c(\rho)$ boundary the secondary maximum occurs close to the midpoint (particularly for large $l$) but rapidly tends to the position of the secondary source.

\section{Repeated observations of a fixed pair of sources}
\label{sec:obs}

If two sources are fixed, in relative orientation and distance, then repeated observations taken at different angles, $\phi$, can mimic apparent astrometric positional offset, and in the case of an elongated PSF, also astrometric motion.

This is relevant for true binary systems, provided they have minimal relative motion (i.e. a period substantially longer than that over which observations are taken). In general blended foreground and background sources will likely have different parallaxes and proper motions and thus vary their distance and orientation on timescales of months. Very distant bright sources, with negligible motion, and closer pairs with fortuitously matching parallax and proper motion, will also have negligible relative motion. This analysis can also apply in any case where the timespan over which the observations were taken was short enough ($\ll 1$ year).

Cases with some relative motion can of course be modelled numerically, but for the rest of this section we focus on the most analytically tractable case of a pair of sources that maintain their relative configuration.

For an elongated PSF we can think of $\alpha$ and $\beta$ as the width of the PSF \textit{along-axis} and \textit{across-axis}  respectively\footnote{For the \textit{Gaia} mission these correspond to the directions parallel and perpendicular to the path the telescope's field of view traces out on the sky}. As we have already shown, the offset of the photocenter (or primary maxima if the brightness profile is multi-peaked), $x_1$, is a function of $r$, $l$ and the projected width of the ellipsoid along the scan-angle, $\gamma(\phi)$.

\begin{figure}
\includegraphics[width=0.98\columnwidth]{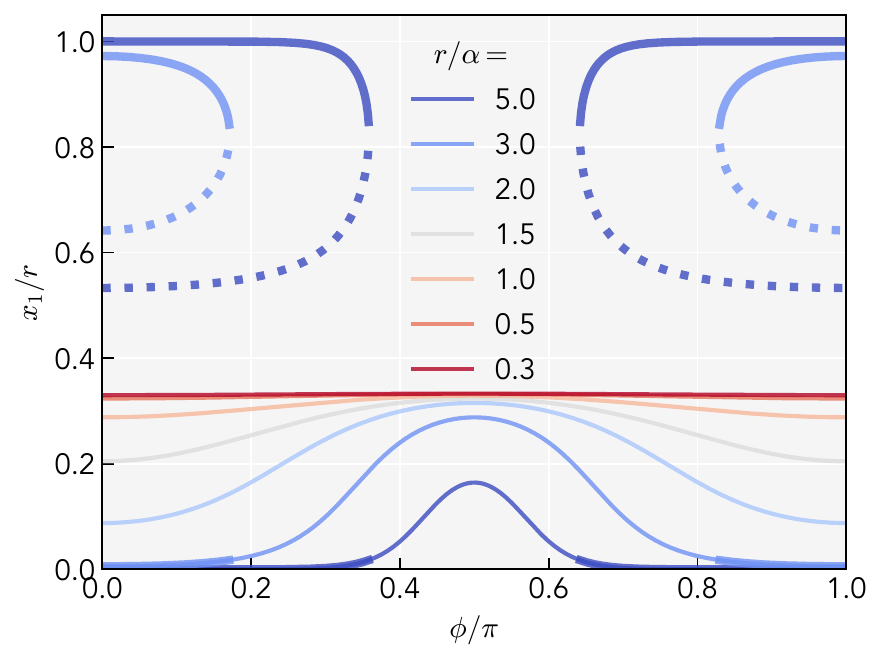}
    \caption{The variation of the position of the observed maxima, as a function of scan angle $\phi$, of a system with $l=\frac{1}{2}$ and a range of $r$. As in figure \ref{fig:col_grid} we use $\beta /\alpha=3$. The two largest separation systems (darkest blues) are resolved at some angles (small $|\sz_{\phi}|$) as shown with a thicker line, and with the position of the minima and secondary maxima also shown as dotted and solid lines respectively. As all expressions involved are even functions of $\phi$ we need only show the behaviour for $0\leq \phi \leq \pi$.}
    \label{fig:col_scan}
\end{figure}

In figure \ref{fig:col_scan} we show the variation of $x_1$ as a function of $\phi$, for an example system with $l=\frac{1}{2}$ and a variety of $r$. We use a ratio of $\beta/\alpha=3$, the approximate elongation of the \textit{Gaia} PSF, throughout the rest of this section. For small separations $x_1$ is roughly constant (tending to $\frac{l}{1+l}r$) but for $r\gtrsim \alpha$ we see visible variation in the photocentre as a function of scan angle. 

The maximum inferred separation occurs when $|\sz_\phi|\rightarrow 1$ and $\gamma$ is close to it's maximum value of $\beta$, hence the two sources are maximally blended. For larger seperations the secondary maxima is resolved for scan angles such that $|\sz_\phi|$ is small (where $\gamma$ is close to it's minimum value of $\alpha$).

A system will be \textit{partially resolved}, appearing as a single source for some scan angles and two sources for others, if
\begin{equation}
\label{eq:scan_resolve}
\alpha  < r/\rho_c(l) < \beta 
\end{equation}
(see section \ref{sec:resolvability}). Systems which are more compact than this will always be unresolved, whilst wider systems will always have two visible brightness peaks.

We can consider the along- and across-scan positions of the photocentre as two separate measurements:
$\delta_{al} = \cz_\phi \cdot x_1$ and $\delta_{ac} = \sz_\phi\cdot x_1$
with associated widths $\alpha$ and $\beta$ respectively.

\begin{figure}
\includegraphics[width=0.98\columnwidth]{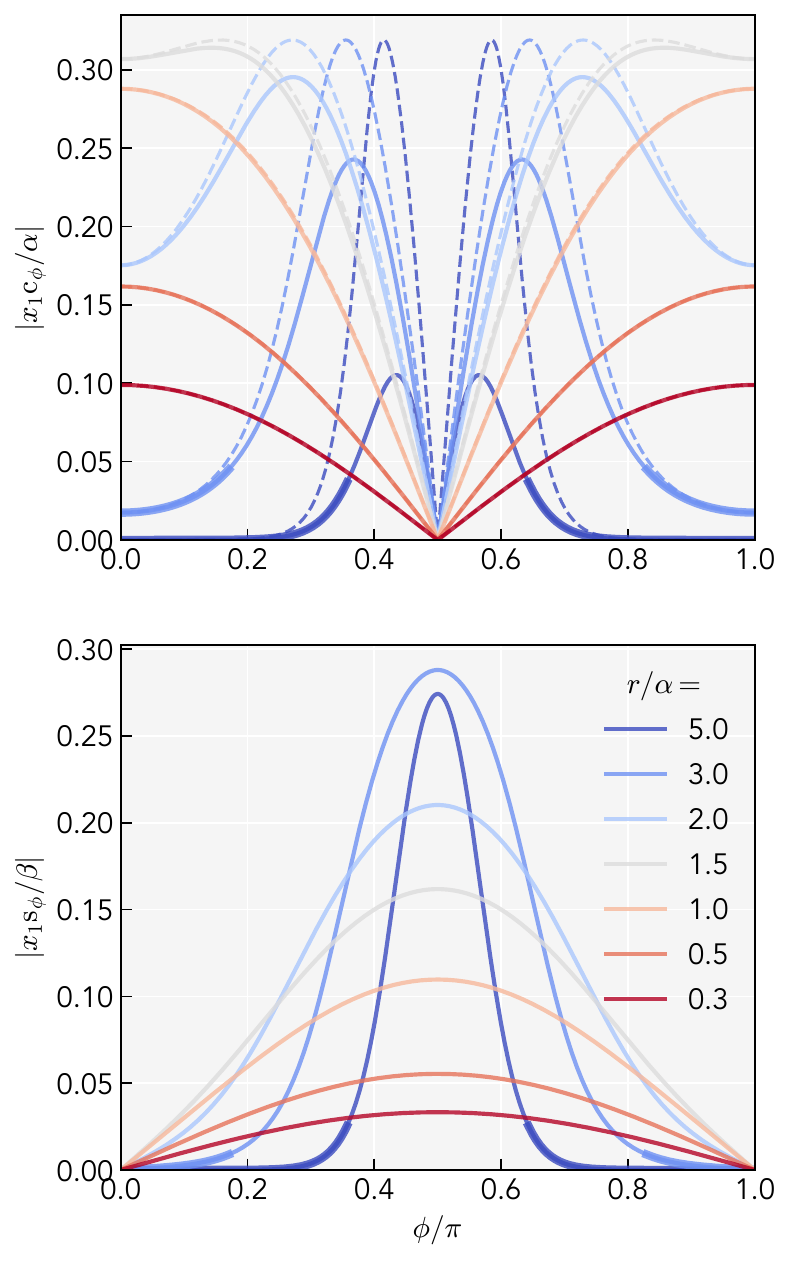}
    \caption{Similar to figure \ref{fig:col_scan} we show the variation of the along-scan (top) and across-scan (bottom) position of the observed maxima, now in units of their relative width. Here we show the absolute value, noting that the actual sign of the observed offset is the same as the sign of the trigonometric term. In the across-scan plot we also show (dashed lines) the offset as would be calculated taking $\beta\rightarrow \infty$.}
    \label{fig:col_scans}
\end{figure}

In figure \ref{fig:col_scans} we show the along- and across-scan positions for the same $l=\frac{1}{2}$ system. Though the photocentre displacement is maximum when $\phi=\pm\frac{\pi}{2}$, this corresponds to the pair lying perpendicular to the scan and hence $\delta_{al}=0$. At small separations $\delta_{al}$ is largest close to the scan direction (small $|\sz_\phi|$) but for more distant sources the asymmetry in $\gamma$ leads to the maximum offset occurring when the pair is close to perpendicular ($|\sz_\phi|$ close to but less than 1). $\delta_{ac}$ is always largest for perpendicular scans.

We also show the across-scan position that one would find if $\beta\rightarrow \infty$ - as is implicitly assumed by the calculation in \citet{Lindegren22}. This would lead to greater predicted offsets, especially at large separations, as any source observed at an angle such that $|\cz_\phi|\lesssim\alpha/r$ would appear to be completely blended.

\subsection{Binary offset and error}

\begin{figure*}
\includegraphics[width=0.98\textwidth]{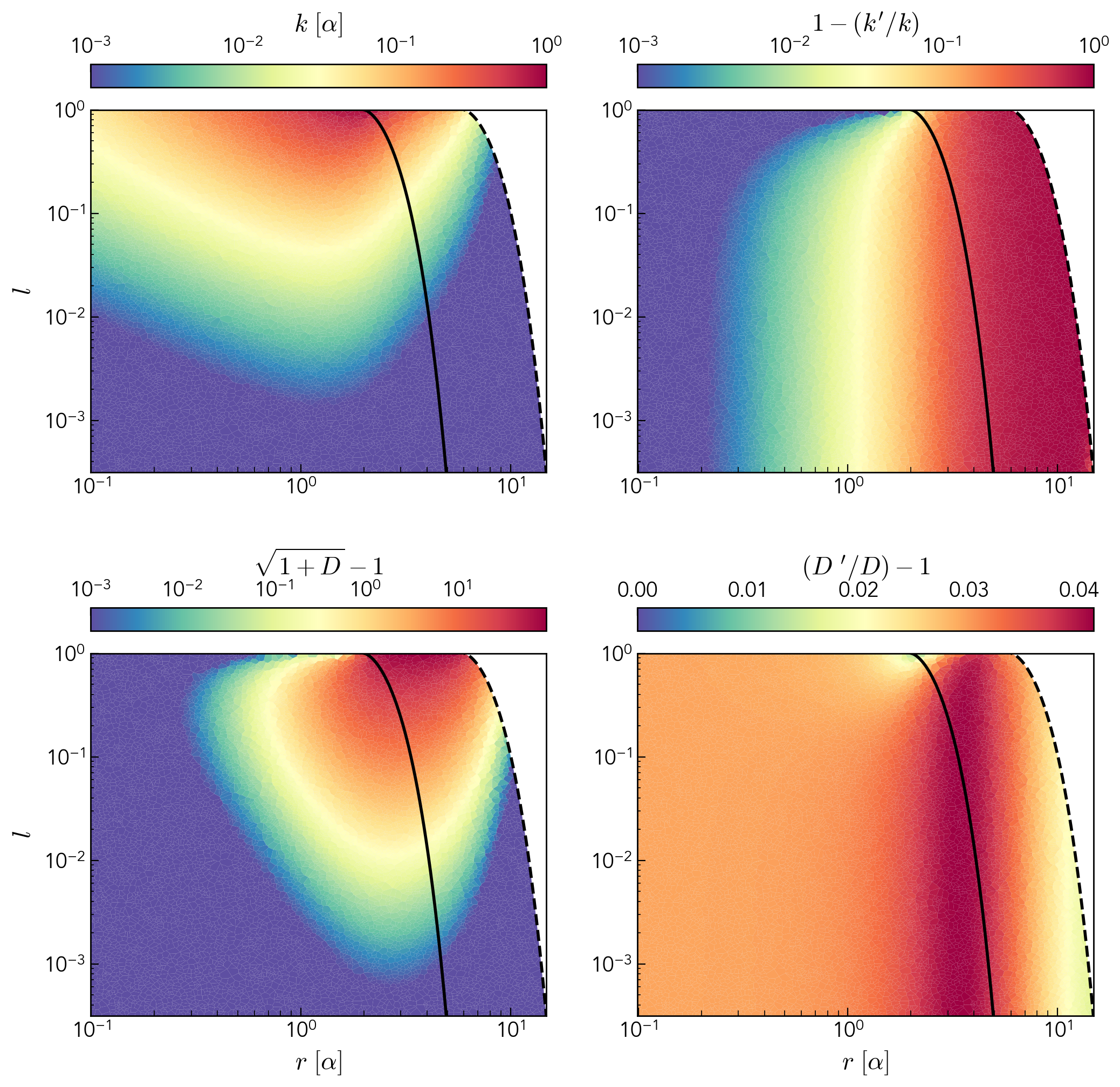}
    \caption{(Left panels) The fitted position offset, $k$, and excess noise, $D$, of a stationary pair of sources (for example a long period binary) observed over a range of scan angles. The excess noise is shown in a form rescaled to match excess RUWE (see text) with significant values being $\gtrsim 0.1$. (Right panels) We compare these to the equivalent fits using only the along-scan measurements, $k'$ and $D'$. $k'$ is always $<k$ and $D'$ always $>D$. To the left of the thick solid line both sources are always unresolved, to the right of the dashed line both are always resolved (and hence we do not perform the fit), and between the lines the sources are unresolved at some angles (see equation \ref{eq:scan_resolve}).}
    \label{fig:col_binary}
\end{figure*}

It has been convenient to work in coordinates ($x,y$) relative to the position of the primary, but if we were to consider a single source at an offset position ($\tilde{x},\tilde{y}$) it would have apparent along- and across-scan displacements of $d_{al}=\tilde{x}\cz_\phi + \tilde{y}\sz_\phi$ and $d_{ac}=-\tilde{x}\sz_\phi + \tilde{y}\cz_\phi$. Inverting this argument, some photocentre shift can be captured as a simple position offset (and thus just a shift of the apparent location of the source). 

We can assume that this shift, much like the photocentre, lies along the line connecting the sources, such that $\tilde{y}=0$. Averaged over $\phi$ there will be some best-fitting $\tilde{x}=k$ that minimises the apparent error in the source position.

We can define the photocentre offsets along and across the scan direction as
\begin{equation}
\Delta_{al}=\cz_{\phi}\cdot(x_1-k) \ \ \mathrm{and} \ \ \Delta_{ac}=\sz_{\phi}\cdot(x_1-k).
\end{equation}

The precision to which the position of a source can be measured depends on the width of the PSF and how well sampled it is. For $N$ samples (i.e. measured photons) of a distribution with PSF width $\sigma$ we might expect a positional uncertainty of $\sim N^{-\frac{1}{2}}\sigma$. Other observational constraints, such as saturation of CCD pixels, can limit the possible precision. \citet{Lindegren21} (specifically appendix A) detail the current along-scan astrometric precision of \textit{Gaia}, showing that it is photon noise dominated for systems with apparent magnitude $m_G\gtrsim 14$, and for brighter sources is approximately constant, with $\sigma_{al}\sim 0.2$ milliarcseconds.

More generally we can define $\mu=\sigma_{al}/\alpha$. For \textit{Gaia} $\mu\sim1/200$ for bright sources and decreases for $m_G\gtrsim 14$. We will assume this ratio is the same for across-scan measurements. 

The total relative offset, normalised by the precision, thus follows
\begin{equation}
\bar{\Delta}^2 = \frac{\Delta_{al}^2}{\sigma_{al}^2} + \frac{\Delta_{ac}^2}{\sigma_{ac}^2} = \frac{(x_1-k)^2}{\mu^2\gamma^2}.
\end{equation}

We denote the average over all $\phi$\footnote{For all quantities of interest in this section the full behaviour is actually captured by any individual quadrant, so we can reduce the cost of numerical integration by only averaging over $0\leq \phi\leq \frac{\pi}{2}$.} as $\langle \rangle$ such that
\begin{equation}
\langle q \rangle = \frac{1}{2\pi}\int_0^{2\pi} q \cdot d\phi.
\end{equation}
If $\phi$ was non-uniformly sampled $\langle q \rangle$ could instead be reimagined as the weighted average over the distribution $p(\phi)$, though we will assume uniformity for the rest of this section.

We can define
\begin{equation}
D(k) = \left\langle \bar{\Delta}^2 \right\rangle,
\end{equation}
the average (normalised) deviance between the photocentre and the offset position.

The value of $k$ that minimises $D$ will obey $\partial_k D=0$ and hence we can find
\begin{equation}
\label{eq:k_full}
k= \left\langle \frac{1}{\gamma^2}\right\rangle^{-1} \left\langle\frac{x_1}{\gamma^2}\right\rangle
\end{equation}
and thus
\begin{equation}
\label{eq:D_full}
D=\frac{1}{\mu^2}\left[\left\langle \frac{x_1^2}{\gamma^2}\right\rangle -\left\langle \frac{1}{\gamma^2}\right\rangle^{-1} \left\langle\frac{x_1}{\gamma^2}\right\rangle^2\right].
\end{equation}

We also note one quirk of \textit{Gaia}, that for all but the brightest sources only the along-scan measurements are included in the astrometric fit. The equivalent best-fit offset, $k'$, thus corresponds to the minimal value of
\begin{equation}
D'(k') = \left\langle \frac{\Delta_{al}^2}{\sigma_{al}^2} \right\rangle.
\end{equation}
which has solution
\begin{equation}
k'= 2 \left\langle x_1 \cz_\phi^2\right\rangle
\end{equation}
and
\begin{equation}
D'=\frac{1}{\mu^2\alpha^2}\left( \left\langle x_1^2 \cz_\phi^2\right\rangle -2 \left\langle x_1 \cz_\phi^2\right\rangle^2 \right).
\end{equation}
Note that while these omit the across-scan measurements, the across-scan error still appears in $\gamma$ and thus affects $x_1$.

\subsection{Observational expectations}

We can consider the relevance of this calculation to a fit performed on a real observational dataset, with associated noise. From a series of observations, at a variety of times and scan angles, it is possible to estimate the position of a source, and also it's motion. 

Assuming a single source (as the main \textit{Gaia} source catalogue does) this motion is a combination of proper motion and parallax. If the scans are frequent and randomly oriented we would not expect any consistent behaviour that could mimic astrometric motion, thus we will only consider the position offset in detail in this work. However if there is some aliasing or coherent time evolution of scan-angles and times the varying positional offset may mimic motion, as seen in \citet{Holl23}.

If there is excess unmodelled astrometric motion, for example from binarity, there will be a poor astrometric fit. Due to the ubiquity of binaries and higher multiples (accounting for around half of all systems \citealt{Raghavan10}) persistently poor fits can be used to identify probable multiple systems (see e.g. \citealt{Penoyre20,Belokurov20,Penoyre22b}).

It is common to quantify the goodness-of-fit with a measure like the reduced chi-squared, $\chi^2_r$, the difference between the observations and a model, normalised by observational error, which we expect to be close to one for a well fitting model. On top of the random noise we would expect our excess noise to inflate the error leading to
\begin{equation}
\label{eq:redchisq}
\chi^2_r \sim 1+D.
\end{equation}

The significance of the excess noise depends on the survey, primarily through the number of observations. We can compare directly to \textit{Gaia} DR3's published Renormalized Unit Weight Error ($RUWE$) values, which record the square root of the $\chi^2_r$ values of the astrometric fit. Using equation \ref{eq:redchisq} we can approximate the excess $RUWE$ as
\begin{equation}
RUWE-1 \sim \sqrt{1+D} - 1
\end{equation}
In $\textit{Gaia}$ DR3 a system can be distinguished as a significantly bad fit (whether due to internal motion or other contaminants) if the excess $RUWE$ is $\gtrsim 0.25$ and thus \citep{Penoyre22,Castro-Ginard24}. 


We should note though that the $D$ value for a given system and instrument is constant, whilst longer observation times can reduce the other sources of noise making $D$ more significant (as will be the case in future \textit{Gaia} data releases, see for example Guerriero et al. in prep.).


\subsubsection{Observable offset and error}

In figure \ref{fig:col_binary} we show the offset, $k$, and deviance, $D$, for a variety of systems of different $r$ and $l$ (always with $\beta/\alpha =3$). We can find $x_1$ equally well in the resolved and unresolved case, though we note that in the partially resolved region (between the black lines) we would also expect to sometimes separately resolve the secondary (and thus have extra evidence that the system is a binary, and of its properties).

We see that the position offset, $k$, is maximal for $l=1$ and $r=2\alpha$ (which also corresponds to the maximum of $x_1$, see \ref{ap:max_offset}). At lower radii ($r\lesssim \alpha$) the offset is well approximated by $x_0=\frac{l}{1+l}r$, a constant with no angular dependence, and hence $D$ is small. Significant offsets are mostly associated with high light ratios, and remain significant even at relatively small separations.

We see the largest deviations from good astrometric fits, $\sqrt{1+D}-1$, at high light ratios and in the partially resolved region. The excess noise can be very large, well above the threshold of 0.25 appropriate for current \textit{Gaia} data. Even excluding the semi-resolved region excesses of up to 50 are possible. This is a relatively hard predicted upper limit, and agrees well with the maximum RUWE seen in sources separated by around 1 arcseond (\citealt{Belokurov20}, figure 5).

In contrast to the offset, the excess noise is significant only for relatively large separations ($r \gtrsim \alpha$) but down to relatively small light ratios ($\gtrsim 10^{-2}$).

This suggests three regimes of blended binaries affecting observed astrometry:
\begin{itemize}
\item \textit{offset and error} - $r\gtrsim \alpha,\  l\gtrsim 0.1$ - sources are significantly offset with high excess noise, which should make them easy to flag as erroneous
\item \textit{offset, no error} - $r\lesssim \alpha,\  l\gtrsim 0.1$ - sources are significantly offset but with negligible associated noise (as in this regime the photocentre offset is not scan angle dependent), thus these would likely be considered single stars though the measured positions would be displaced from their true value
\item \textit{error, no offset} - $r\gtrsim \alpha,\  0.1\gtrsim l\gtrsim 0.01$ - the observed position of these sources is essentially correct, but they still have large errors associated with the scan-angle dependence of the elongated PSF
\end{itemize}
Outside of these regimes a static binary has little impact on the measured astrometry.

\subsubsection{Observational complications}

Systems in the \textit{offset and error} regime are the most amenable to detection with other methods, more observations or in later surveys. The large light ratios make them easier to discern from single sources, either from elongated photometry, or two discernable profiles in their spectra. \textit{Gaia} includes flags based on the Image Parameter Determination (IPD) pipeline that record deviations from expected point-source brightness profiles which can be used to flag some sources in this regime (see for example appendix B of \citealt{Penoyre22b}). The pipeline for \textit{Gaia} DR4 will identify and fit secondary peaks where possible, though the regime of interest here is at the edge of, or below, the threshold of resolvability and thus this will mitigate the effects mostly of the partially resolved systems.

One caveat here is background noise, both instrumental error and the integrated light from distant stars. In general if the brightness of the secondary drops below this background the effect should be diminished if not entirely negated.

Another concern for \textit{Gaia} observations is that these large astrometric deviations may interfere with the assignment of individual measurements to a given source. This is done by spatial clustering over a scale of typically 0.1 arcsecond, or about 2.5$\alpha$ (see for example \citealt{Fabricius16,Torra21}). The photocenter offset should never be more than $\alpha$, suggesting that the pipeline should be relatively robust, but some semi-resolved systems are approaching the limiting threshold and thus we may expect occasional contamination/ lost measurements. This problem compounds in dense fields, where multiple sources may be blended in some measurements and recorded separately at other times.


\subsubsection{Along-scan only measurements}

We also compute $k'$ and $D'$, the equivalent measures when only along-scan measurements are used in the fit. $k'$ is always an underestimate, as we are leaving out the extra contribution of the across-scan motion. The difference is small for $r<2 \alpha$ but by $r=4\alpha$ it is approximately a factor of 2. Conversely $D'$ is always a slight overestimate, due to ignoring the across-scan noise $\beta$. This is true even at low $r$ where the inflation factor tends to constant $\sim2-3\%$. $D'/D$ is largest at $3<r/\alpha<5$ and is smallest close to $r=2\alpha,\ l=1$.

Depending on the use case the degree of agreement between $k$ and $k'$, and $D$ and $D'$, may be either heartening or troubling. The measured offset is accurate for essentially all $r\lesssim \alpha$ binaries, and whilst the overestimate of the noise is persistent, it is small (of order a few percent). However if one's focus was partially resolved binaries ($r\gtrsim \alpha$) we can see that we are losing significant information when restricted to across-scan data only, especially about the true offset (and hence the inferred separation and light ratio).

\section{Conclusions}

In this work we have derived analytic expressions for the positions of maximal brightness of blended sources, the conditions under which one or both sources would be resolvable, and the impact of consistently blended sources on repeated astrometric observations.

We start, in section \ref{sec:coord}, with an elongated Gaussian PSF, as is appropriate for surveys such as the \textit{Gaia} mission, with a rectangular focussing mirror. We show that this more general case can be reduced to a 1D problem, and thus converges with the circular PSF case when normalised by a scan-angle dependant effective PSF width, $\gamma(\phi)$ (equation \ref{eq:gamma}). Thus any general case can be expressed in terms of $\rho=\frac{r}{\gamma}$, where $r$ is the on-sky separation of the sources. This introduces a further dependence on the relative orientation of the sources and our instrument, and shows that when this orientation changes over multiple measurements the variation in observed photocentre can provide extra information about the relative separation and light ratio of the sources. Such variations have indeed been observed in \textit{Gaia} already, as detailed in \citet{Holl23}.

Having reduced the general case to a simpler 1D case we then find, in section \ref{sec:resolvability}, the conditions for which their are two distinct maxima in the brightness profile (which is approximately equivalent to the condition that the two sources can be distinctly observed). We find an analytic form of the critical light ratio, $l_c$ (equation \ref{eq:l_c}), below which a source of a given separation will only have a single brightness maximum. This can be inverted numerically into a limiting separation for a given light-ratio, $\rho_c(l)$, below which the two sources would be blended, the more directly useful case for objects with constant luminosity moving relative to each other on-sky.

The positions of the primary, and where they exist secondary, maxima (which we associate with the positions that would be measured by an instrument observing the system) can be found numerically and we provide recipes for doing so in section \ref{sec:numerical_solutions}.

When two blended sources maintain a fixed orientation analytic expressions for their position offset and induced error can be derived, as we do in section \ref{sec:obs}. This is a particular case of interest for long period (and hence effectively stationary) binaries observed at various times and scan-angles. If the separation is small the blending of the sources corresponds to a simple shift in the inferred position. As we approach large separations ($\rho \gtrsim 1$) the asymmetry of the effective width $\gamma$ introduces extra angular dependence which cannot be captured by a simple offset and leads to excess noise on astrometric fits. Depending on the pattern of scan angles with time this noise may translate into a bias in proper motion or parallax (for further discussion see \citealt{El-Badry24b}). Blends between unrelated background and foreground stars will show the same behaviour, provided that their astrometric motion (parallax and proper motion) are similar or negligible.

We show that pairs which are close to being resolved cause a persistent and potentially large inflation to the inferred error on an astrometric fit, even at small light ratios which would likely make them otherwise undetectable. This can increase the reduced chi squared of such a fit by enough to flag them as a bad single-body astrometric fit, and thus infer that they are likely a binary/blend. 

Thus a population of long period semi-resolved binaries also contributes to the known population of binaries with significant internal astrometric motion (periods of order of the timespan of the observations, of order years for $\textit{Gaia}$ \citealt{Penoyre22}). With epoch astrometry data (as will be available in the next \textit{Gaia} data release) these two cases can be differentiated, but currently they cannot easily be separated, even though they represent markedly different types of binary.

Fitting astrometry to only the 1D along-scan positions (as \textit{Gaia} does for dimmer sources) inflates the noise contribution by a few $\%$ for almost all systems, with the largest effect for partially resolved pairs.

In appendix \ref{ap:max_offset} we find the conditions which maximise the photocentre offset of the primary peak - i.e. which give the largest deviation between the measured position of the blended source and the true location of the primary.

Finally in appendix \ref{ap:approx} we extend the zeroth order solution for the position of the photocenter, $x_0$, to the next order terms of order $\rho^2$. We find the positional offset and associated excess noise in this regime, showing that the noise contribution of fixed orientation blended sources can be maximised (or minimised) depending on the degree of elongation of the PSF.

The assumption of a smooth Gaussian PSF, which makes much of what is discussed here analytically tractable, is obviously a major simplification. Real PSFs vary between sources and can change over time with degradation and repairs to the instrument. A real PSF can contain secondary maxima, asymmetries, and extra components not easily predicted from the instrument design. The image is projected onto a finite number of CCDs resulting in pixelated images. A deformed PSF will likely still obey most of the relations presented here, but with a slight (potentially scan angle dependant) rescaling compared to the Gaussian case. Secondary maxima are a more significant challenge, as these can mimic a second source, or overlap with one, and require detailed PSF modelling.


For sources moving relative to each other it is often possible to predict, from photometric or astrometric time series, when they will be blended. Even simple recipes, such as could be made from this analysis, to account for this blending can significantly improve the accuracy of the inferred brightness and motion of the brighter source, and recover information about the dimmer source that would otherwise be lost in the blend.

Sources which remain close over the full timespan of observations are more difficult to unpick, although as we have shown may be detectable through excess noise and other deviations from ideal astrometry.

Blended sources are both a ubiquitous nuisance signal for observing single objects, and a common prospect for observing binaries (or higher multiples). They distort the apparent motion and cause changes in the inferred brightness and characteristics of a source. But if one or both sources is well characterised the blending can provide new insight on the presence of un- or under-resolved objects. Thus this fuller depiction of the properties and behaviours of blended point sources may be of relevance and utility for any stellar survey, and for better separating physical behaviours from quirks of the projection along which we view them.

\section*{Data Availability}

The data underlying this article will be shared on reasonable request to the corresponding author.

\section*{Conflicts of interest}

We are not aware of any conflicts of interest associated with this work.

\section*{Acknowledgements}
We would like to thank Berry Holl and Anthony Brown for the discussions and input that led to this piece of work. We would also like to thank Emily Sandford, Michael Davidson, Hugh Osborn and the anonymous reviewer for comments and improvements to the paper. ZP acknowledges
support from European Research Council (ERC) grant number:
101002511/project acronym: VEGA P.

\bibliographystyle{mnras}
\bibliography{bib}
\bsp

\appendix

\section{Maximal photocentre offsets}
\label{ap:max_offset}

In this appendix we are interested in which $\rho$ ($=r/\gamma(\phi)$) and $l$ give the largest photocentre offsets - i.e. which blended sources will appear most displaced. 

We will assume only one maximum is present (i.e. $l<l_c(\rho)$, equation \ref{eq:l_c}) and thus a single solution, $x_m=x_1$, for equation \ref{eq:x0}. We want to maximise $x_m$, or equivalently $\chi_m$ ($=x_m/\gamma$) which satisfies
\begin{equation}
\label{eq:ap_chi0}
\chi_m = \frac{l \rho}{l+\lambda_m}
\end{equation}
where
\begin{equation}
\label{eq:ap_lambda0}
\lambda_m = e^{\frac{\rho}{2}(\rho - 2\chi_m)}.
\end{equation}

We have already shown that the separation, $x_1$, has a maximum value $\rightarrow x_0$ as $\rho\rightarrow 0$ (figure \ref{fig:col_x}). Here we derive the maximum absolute separation.

Differentiating equation \ref{eq:ap_chi0} we have
\begin{equation}
\partial_\rho \chi_m = \frac{l}{(l+\lambda_m)^2}\left(l+\lambda_m - \rho \ \partial_\rho \lambda_m \right)
\end{equation}
and
\begin{equation}
\partial_l \chi_m = \frac{\rho}{(l+\lambda_m)^2}\left(\lambda_m - l \ \partial_l \lambda_m \right).
\end{equation}

Differentiating equation \ref{eq:ap_lambda0}
\begin{equation}
\partial_\rho \lambda_m = \left(\frac{\rho \lambda_m}{l+\lambda_m} -\rho \ \partial_\rho \chi_m\right)\lambda_m
\end{equation}
and
\begin{equation}
\partial_l \lambda_m = -\rho \lambda_m \ \partial_l \chi_m.
\end{equation}

Putting these together and rearranging we find
\begin{equation}
\label{eq:d_rho_chi}
\partial_\rho \chi_m = \frac{l}{l+\lambda_m}\frac{(l+\lambda_m)^2 - \lambda_m^2 \rho^2}{(l+\lambda_m)^2 - l \lambda_m \rho^2}
\end{equation}
and
\begin{equation}
\label{eq:d_lambda_chi}
\partial_l \chi_m = \frac{\lambda_m \rho}{(l+\lambda_m)^2 - l \lambda_m \rho^2}.
\end{equation}

The common denominator of equation \ref{eq:d_rho_chi} and \ref{eq:d_lambda_chi} is always positive. Thus $x_m$ is maximised for $l=1$ (as $\partial_l \chi_m > 0$ always).

Let us denote the maximal $x_m$ for a given $l$ as $x_{max}(l)$ (with corresponding $\chi_{max}$, $\lambda_{max}$ and $\rho_{max}$). Setting $\partial_\rho \chi_0|_{x_{max}}=0$ we find
\begin{equation}
\label{eq:lambda_max}
\lambda_{max}=\frac{l}{\rho_{max} -1}
\end{equation}
(discarding a negative root) and putting this into equation \ref{eq:ap_chi0} we find
\begin{equation}
\chi_{max}=\rho_{max} -1 = \frac{l}{\lambda_{max}}.
\end{equation}
Equation \ref{eq:lambda_max} can be combined with equation \ref{eq:ap_lambda0} and solved numerically. In the limit of small $l$, $\rho_{max}\rightarrow 1+l e^{-\frac{1}{2}}$, and thus $\chi_{max} \rightarrow l e^{-\frac{1}{2}}$.

\begin{figure}
\includegraphics[width=0.98\columnwidth]{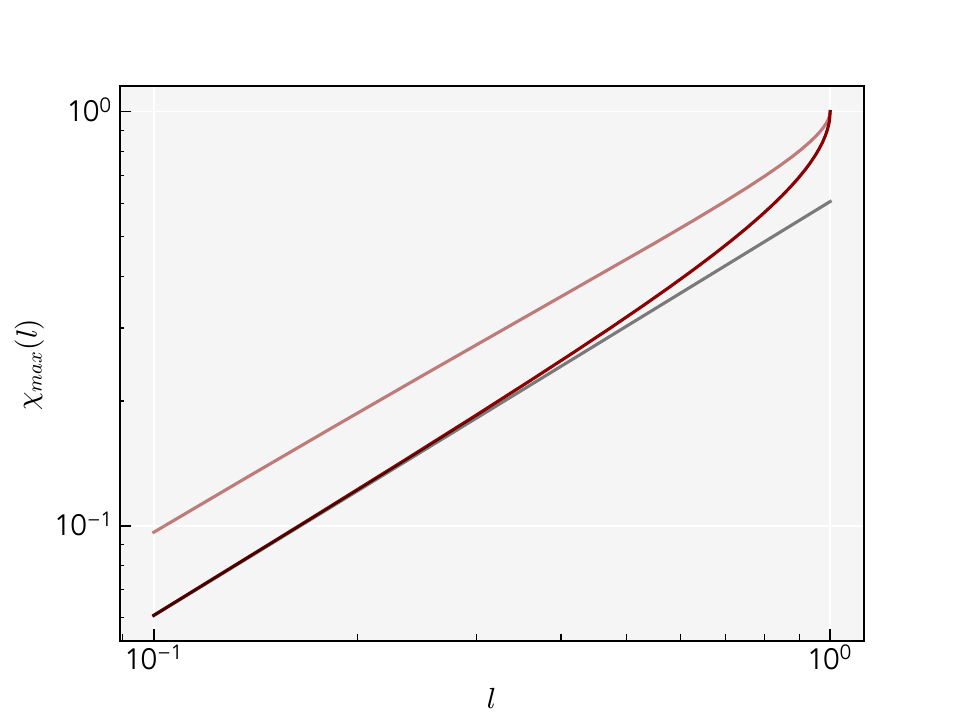}
    \caption{Solid line: maximum possible displacement of the primary peak, $\chi_{max}=x_{max}/\gamma$, as a function of $l$. Dashed line: the overestimated value that would be derived from assuming $\chi=\rho l /(1+l)$. Dotted line: the limiting value at low $l$ of $l e^{-\frac{1}{2}}$.}
    \label{fig:x_max}
\end{figure}

We can solve this to find $x_{max}(l)$ as shown in figure \ref{fig:x_max} (in natural units of $\gamma$). As expected it is maximal for $l=1$, where $\chi_{max}=1$. The solution drops off faster than the idealised overestimation, of $x_0=\rho l /(1+l)$ and quickly tends to the limiting value of $l e^{-\frac{1}{2}}$, becoming visually indistinguishable for $l\lesssim \frac{1}{3}$.

\section{Approximate form for low $\rho$}
\label{ap:approx}

As $\rho=\frac{r}{\gamma} \rightarrow 0$, $\lambda\rightarrow 1$ and the single zeroth order approximate solution to equation \ref{eq:x0} is $x_m \approx x_0=\frac{l r}{1+l}$. In this appendix we wish to find the next term in an expansion for low $\rho$.

We start with the form
\begin{equation}
\label{eq:x0_ap}
x_m = \frac{r l}{\lambda_m + l} = \frac{r}{2}\left( 1-\frac{2 \ln \lambda_m}{\rho^2}\right)
\end{equation}
where the latter equality comes from rearranging equation \ref{eq:chi_lambda}). 

Assuming $\rho<2$ (which we will see later must be the case for this low-order expansion to be accurate) there is only maximum, near the primary source at $x_m=x_1 \leq \frac{r}{2}$ and thus $\lambda_m
\geq 1$. We will make the assumption that $\lambda_m-1 \ll 1$ and hence define
\begin{equation}
\ln \lambda_m = \epsilon \ll 1.
\end{equation}
Similarly expanding $\lambda_m = e^\epsilon = 1+\epsilon+\mathcal{O}(\epsilon^2)$ and substituting both into equation \ref{eq:x0_ap} we find
\begin{equation}
1-\frac{2 \epsilon}{\rho^2}=\frac{2 l}{l+1+\epsilon+\mathcal{O}(\epsilon^2)}=\frac{2l}{1+l}\left(1-\frac{\epsilon}{1+l} +\mathcal{O}(\epsilon^2)\right)
\end{equation}
and rearranging we find
\begin{equation}
\epsilon +\mathcal{O}(\epsilon^2) =\frac{\rho^2}{2}\frac{1-l}{1+l}\left( 1-\frac{l\rho^2}{(1+l)^2}\right)^{-1}.
\end{equation}
From this we can see that the requirement that $\epsilon \ll 1$ is equivalent to $\rho^2 \ll 1$ and thus we can further simplify to
\begin{equation}
\epsilon = \frac{\rho^2}{2}\frac{1-l}{1+l}\left( 1+\frac{l\rho^2}{(1+l)^2}\right) +\mathcal{O}(\rho^4).
\end{equation}
Substituting this into the second equality of equation \ref{eq:x0_ap} we get
\begin{equation}
x_m = \frac{lr}{1+l}\left(1-\frac{l}{(1+l)^2}\rho^2 \right) +\mathcal{O}(\rho^4) 
\end{equation}
which can be written in terms of $x_0$ as 
\begin{equation}
\label{eq:xm_simple}
x_m \approx x_0\left(1- \frac{x_0^2}{l \gamma^2}\right)
\end{equation}
(note that as $l\rightarrow 0$, $x_0\rightarrow l$ and hence right hand term $\rightarrow0$).

We learn two significant things from this: that the next order correction to $x=x_0$ is $\mathcal{O}(\rho^2)$,
and that this is the lowest order at which $\gamma$, the orientation dependant effective width, affects the solution.

\subsection{Offset and error}

We can repeat the analysis of section \ref{sec:obs} to find the observed positional offset and associated error for this second order solution.

Equations \ref{eq:k_full} and \ref{eq:D_full} can now be written as
\begin{equation}
k=x_0 - \frac{x_0^3}{l}\frac{\left\langle\gamma^{-4}\right\rangle}{\left\langle\gamma^{-2}\right\rangle} +\mathcal{O}(\rho^4) 
\end{equation}
and 
\begin{equation}
D=\frac{x_0^6}{l^2}\left(\left\langle\gamma^{-6}\right\rangle -\frac{\left\langle\gamma^{-4}\right\rangle^2}{\left\langle\gamma^{-2}\right\rangle}\right)+\mathcal{O}(\rho^8)
\end{equation}
(where we note that the latter equation remains true when we include the terms of order $\rho^4$ in $x_m$).

Written in terms of $\alpha$ and $\eta=\frac{\alpha^2}{\beta^2}\ (\leq 1)$ we find
\begin{equation}
\left\langle\gamma^{-2}\right\rangle = \frac{1}{2 \alpha^2}(1+\eta),
\end{equation}
\begin{equation}
\left\langle\gamma^{-4}\right\rangle = \frac{1}{8 \alpha^4}(3+2\eta+3\eta^2),
\end{equation}
and
\begin{equation}
\left\langle\gamma^{-6}\right\rangle = \frac{1}{16 \alpha^6}(5+3\eta+3\eta^2+5 \eta^3).
\end{equation}

Thus, in terms of $\alpha$ and $\eta$
\begin{equation}
k=x_0 -\frac{x_0^3}{4l\alpha^2}\frac{3+2\eta +3\eta^2}{1+\eta} +\mathcal{O}(\rho^4) 
\end{equation}
and
\begin{equation}
D=\frac{x_0^6}{32l^2\alpha^6}\frac{1+4\eta-10\eta^2+4\eta^3 +\eta^4}{1+\eta} +\mathcal{O}(\rho^8). 
\end{equation}

For a circular PSF, $\eta=1$, $k$ goes to its minimum value of $x_m$ (as given by equation \ref{eq:xm_simple}) and $D=0$ as here the image has no angular dependence. Reducing $\eta$, starting from a value of 1, $k$ and $D$ increase, with both reaching their limiting value at $\eta=\frac{2}{\sqrt{3}}-1 \approx 0.1547$ where $D \approx 0.03775\frac{x_0^6}{l^2\alpha^6}$. The existence of a $\eta$ which maximises $D$ suggests the possibility of an optimal instrument design for detecting the presence of sub-resolution binaries in images (or alternatively a worst-case design if one wanted to avoid the excess noise that such sources cause). As one final interesting serendipity, the \textit{Gaia} mission design, with $\eta\approx\frac{1}{3}$ gives almost identical $k$ and $D$ as the $\eta=0$ case (though this ceases to be true outside of the $\rho\ll1$ case).

\label{lastpage}

\end{document}